\documentclass[sigconf,nonacm]{acmart}

\AtBeginDocument{%
  }

\copyrightyear{2026}
\acmYear{2026}
\setcopyright{cc}
\setcctype{by-nc-nd}
\acmConference[SIGCSE TS 2026]{Proceedings of the 57th ACM Technical Symposium on Computer Science Education V.2}{February 18--21, 2026}{St. Louis, MO, USA}
\acmBooktitle{Proceedings of the 57th ACM Technical Symposium on Computer Science Education V.2 (SIGCSE TS 2026), February 18--21, 2026, St. Louis, MO, USA}
\acmPrice{}
\acmDOI{10.1145/3770761.3777146}
\acmISBN{979-8-4007-2255-4/2026/02}

\usepackage{tikz}
\usepackage{wheelchart}

\usetikzlibrary{shapes.geometric, arrows, 3d}

\tikzstyle{block} = [rectangle, rounded corners, minimum width=3cm, minimum height=1cm, text centered, draw=black, fill=blue!30]
\tikzstyle{arrow} = [thick,->,>=stealth]

\usepackage[utf8]{inputenc}
\usepackage{pgfplots}

\usepgfplotslibrary{groupplots,dateplot}
\usetikzlibrary{patterns,shapes.arrows}
\pgfplotsset{compat=newest}

\usepackage{comment}
 
\includecomment{commentCDCtrim}
\excludecomment{commentCDCtrim}

\includecomment{addclearpage}
\excludecomment{addclearpage}

\usepackage{cleveref}
\begin{document}

\title{Ungraded Assignments in Introductory Computing: A Report}


\author{Yehya Sleiman Tellawi}
\email{yehyas2@illinois.edu}
\affiliation{%
  \institution{University of Illinois Urbana-Champaign}
  \city{Urbana}
  \state{Illinois}
  \country{USA}
}

\author{Abhishek K. Umrawal}
\email{aumrawal@illinois.edu}
\affiliation{%
  \institution{University of Illinois Urbana-Champaign}
  \city{Urbana}
  \state{Illinois}
  \country{USA}
}

\renewcommand{\shortauthors}{Yehya Sleiman Tellawi and Abhishek K. Umrawal}

\begin{abstract}
This experience report explores the effects of ungraded assignments on the learning experience of students in an introductory computing course. Our study examines the impact of ungraded assignments on student engagement, understanding, and overall academic performance. We developed and administered new ungraded assignments for a required course in the first year of the Computer Engineering curriculum called ECE 120 Introduction to Computing. To assess the effectiveness of our ungraded assignments, we employed a mixed-methods approach, including surveys, interviews, and performance analysis. Our analysis shows a positive relationship between participation in ungraded assignments and overall course performance, suggesting these assignments may appeal to high-achieving students and/or support better outcomes.
\end{abstract}



\maketitle

\section{Problem and Motivation}
Assessment remains central to education, with grades serving as the dominant measure of performance and progress. In Computer Science (CS) and Computer Engineering (CompE), assignments are often evaluated through automated scripts that emphasize correctness, offering limited formative feedback on conceptual understanding. While grades provide accountability, research shows that grade-focused learning can foster anxiety \citep{Nathaniel2018}, reduce intrinsic motivation \citep{Cain2022, Deci2001}, and encourage `bulimic learning,' where students memorize material for exams and quickly forget it \citep{Bensley1992}. These effects are particularly concerning in technical disciplines, where long-term problem-solving skills and conceptual mastery are essential.

To address these challenges, educators have explored alternative strategies that shift the focus from evaluation to learning \citep{Williams2020, Koehler2022, Lingwall2023}. Ungraded assignments represent one such approach: they encourage engagement with course material without the pressure of formal assessment, allowing students to reflect, experiment, and take intellectual risks. Formats may include design exercises, optimization challenges, proofs, reflections, or multimedia tasks. Prior studies suggest that ungraded work can promote collaboration and deeper engagement \citep{Hasinoff2024}, but most of this research has been situated in non-technical fields. This gap motivates our investigation of ungraded assignments in a large introductory computer engineering course. Specifically, we examine how students engaged with optional ungraded tasks, how participation related to academic performance, and what assignment formats were most effective in fostering meaningful learning. We note that, unlike pedagogical `ungrading,' our study focuses on optional tasks that are not included in the final course grade.

\section{Background and Related Work}
Prior work suggests that replacing grades with formative feedback can enhance engagement and learning. Williams \citep{Williams2020} found ungrading boosted participation and performance, while Koehler \citep{Koehler2022} reported gains in reflection and shared decision-making. In engineering, Lingwall \citep{Lingwall2023} showed that ungraded classrooms improved attitudes and exam scores. These studies highlight the promise of ungraded approaches but leave open questions about their effectiveness in large, technical courses such as computer engineering. From a theoretical perspective, self-determination theory \citep{RyanDeci2000} posits that autonomy, competence, and relatedness drive intrinsic motivation. Ungraded assignments primarily support autonomy—students choose whether to complete them—and can foster relatedness through collaborative work unconstrained by grading.

\begin{figure}[h]
    \centering
    \includegraphics[width=0.6\linewidth]{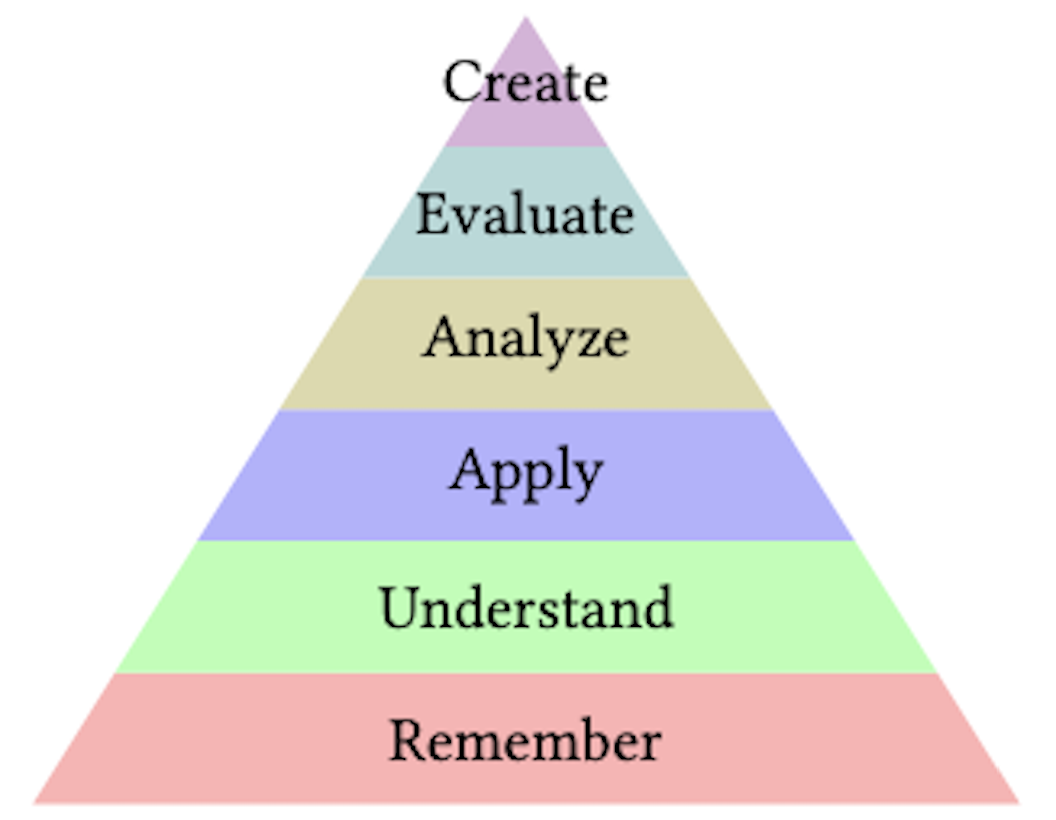}
    \vspace{.01in}
     \caption{Bloom’s Taxonomy \cite{Bloom1956}.}
    \label{fig:bloom}
\end{figure}

Bloom’s taxonomy \citep{Bloom1956} (Figure~\ref{fig:bloom}), which classifies cognitive objectives into six levels—Remember, Understand, Apply, Analyze, Evaluate, and Create—offers a useful lens for ungraded work. While traditional CS and CompE assessments engage higher-order skills, they often prioritize correctness and efficiency over reflection and deeper learning. In this study, ungraded assignments were designed to target advanced cognition: design and proof tasks encourage evaluation and application, optimization challenges promote creativity, and reflective or video-based activities support conceptual understanding. We examine whether such activities can foster learning in a large introductory computer engineering course.

\section{Approach and Uniqueness}
We implemented a series of ungraded assignments in ECE 120: Introduction to Computing, a core course with 336 students in Spring 2025. Students were still provided with detailed feedback by course staff. The assignments were designed to complement graded work by targeting a range of cognitive processes through five main formats (the Bloom's taxonomy category is provided in brackets): 1) Proof-based tasks---reinforcing theoretical concepts (e.g., two’s complement, latches) [Evaluate + Analyze], 2) Design problems---prompting students to apply classroom knowledge to practical trade-offs [Evaluate + Apply], 3)
Optimization challenges---where students minimized LC3 \citep{PattPatel} assembly code length under strict constraints [Create], 4) Reflections---encouraging self-assessment of learning progress [Remember + Understand], and
5) Video-based activities---providing additional context for technical topics like floating-point and error correction [Understand]. Assignments were timed to align with lectures, with lighter workloads during exam weeks. To promote engagement, some tasks incorporated gamification: students completing difficult optimization challenges received recognition on Discord and during review sessions. Evaluation combined quantitative and qualitative methods. Participation data came from Gradescope submissions, and student exam performance was compared between participants and non-participants. To capture perceptions, we distributed an anonymous survey (Likert-type and open-response items) and conducted follow-up interviews on the popularity of optimization challenges relative to other activities.

Our approach is novel in two respects. First, it examines ungraded work in a large technical course, where such interventions are rare compared to humanities or small-class settings. Second, it compares multiple assignment types—ranging from low-stakes videos to intensive optimization problems—while integrating gamified recognition. This combination lets us explore not only whether ungraded assignments support learning, but also which designs most effectively engage students in demanding STEM disciplines.

\section{Results and Contributions}
\begin{figure}[h!]
\centering
\begin{tikzpicture}
    \begin{axis}[
        ymax = 15,
        ybar,
        bar width=0.5cm,
        width=\columnwidth,
        height=0.4\columnwidth,
        enlarge x limits=0.2,
        enlarge y limits=0.1, 
        ylabel={Student count},
        symbolic x coords={
            Two's Complement Proof,
            Logical Completeness Proof,
            Bonus Hardware Lab - Design,
            Simple LC3 Optimization,
            Complex LC3 Optimization,
            LC3 Logic Optimization
        },
        xtick=data,
        nodes near coords,
        x tick label style={
            rotate=90,
            anchor=east,
            font=\small,
            text width= 2.0cm,  
            align=center
        },
        grid=major,
    ]
    \addplot coordinates {
        (Two's Complement Proof, 5)
        (Logical Completeness Proof, 2)
        (Bonus Hardware Lab - Design, 2)
        (Simple LC3 Optimization, 13)
        (Complex LC3 Optimization, 8)
        (LC3 Logic Optimization, 6)
    };
    \end{axis}
\end{tikzpicture}
\caption{Participation counts for each ungraded assignment. Assignments with no student participation are not shown.}
\label{fig:count}
\end{figure}
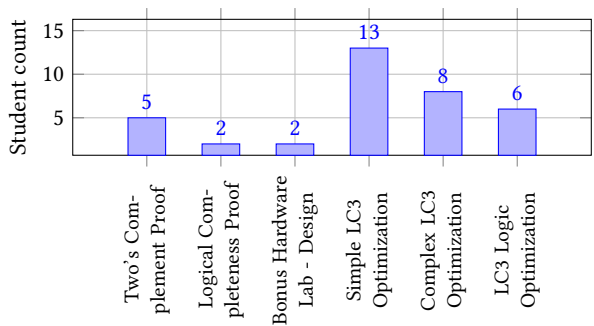

Of the 336 students enrolled in ECE 120, 49 (14.6\%) joined the ungraded assignments' Gradescope, and only 20 students (6\%) completed at least one assignment. Engagement varied considerably across assignment types. LC3 optimization challenges were the most popular: 13 students completed the simple optimization task, 8 completed the complex optimization, and 6 completed the LC3 logic optimization. In contrast, participation in other assignments was significantly lower, with only 5 students attempting the Two’s Complement proof and 2 students each completing the Logical Completeness proof and the Bonus Hardware Lab design task. Assignments with zero participation were excluded from this analysis. Refer to Figure \ref{fig:count}.

\begin{table}[hbt!]
\centering
\setlength{\tabcolsep}{3pt}
\begin{tabular}{l c c c}
    \hline
    \textbf{Exam} & \shortstack{\textbf{Participants}\\\textbf{(Mean $\pm$ SD)}} & \shortstack{\textbf{Non-participants}\\\textbf{(Mean $\pm$ SD)}} & \shortstack{\textbf{Welch's $t$-test}\\\textbf{($p$-value)}} \\
    \hline
    Midterm 1 & 93.8 $\pm$ 5.6  & 83.2 $\pm$ 13.0 & 9.46$\times$10$^{-8}$ \\
    Midterm 2 & 88.6 $\pm$ 9.6  & 73.8 $\pm$ 17.8 & 3.28$\times$10$^{-6}$ \\
    Midterm 3 & 95.2 $\pm$ 4.7  & 80.7 $\pm$ 16.4 & 2.83$\times$10$^{-13}$ \\
    Final     & 91.4 $\pm$ 7.3  & 78.8 $\pm$ 17.2 & 5.50$\times$10$^{-7}$ \\
    \hline
\end{tabular}
\caption{Performance comparison between participants and non-participants in ungraded assignment activities.}
\label{tab:quantitative-results}
\end{table}

Exam performance analysis (Table \ref{tab:quantitative-results}) showed that participants consistently scored higher than non-participants across all exams, with significant differences of 10.6–14.8 percentage points (each $p$-value < 10$^{-5}$). This suggests a positive relationship between participation in ungraded assignments and course outcomes. However, these differences may reflect pre-existing motivation rather than assignments’ effect, as some exams covered unrelated content.

Likert-type survey responses indicated high perceived difficulty and strong valuation of extra credit as a motivator. Time investment and perceived usefulness varied with assignment type, reflecting the diversity of tasks from short videos to challenging optimization problems. Open-response feedback highlighted several themes: students appreciated technical and creative challenges, desired earlier promotion of assignments, and often required extrinsic incentives to engage. Some students reported partially completing tasks without submission, often due to perceived effort or perfectionism concerns. Interviews reinforced that framing assignments as `challenges' and integrating them with graded work increased engagement.

These findings suggest several contributions. First, ungraded assignments may attract highly motivated students and provide opportunities for deeper engagement with higher-order cognitive tasks. Second, gamified or creative challenges (e.g., LC3 optimization) enhance participation, highlighting the importance of task framing. Third, participation and course performance are positively related, indicating that ungraded assignments may appeal to high-achieving students or potentially support improved outcomes. Collectively, these insights guide the design of optional, ungraded interventions in technical courses, emphasizing challenge, creativity, and incentives to maximize impact.

\section{Conclusion and Future Work}
This study found that although participation in ungraded assignments was limited, students who engaged were self-motivated and reported meaningful learning benefits. Well-designed ungraded tasks—especially those emphasizing creativity and challenge—can enhance engagement beyond traditional coursework. 

For the future offerings, visibility is key. Regular mentions in lectures, discussions, and office hours, along with posting assignments on the main course page, would better integrate them into students’ workflow. Repeating the study in the fall, when most computer engineering majors take ECE 120, may yield insights into how student background and major affect engagement.

\section{Acknowledgments}
This research was supported by the Grants for Advancement of Teaching in Engineering (GATE) Program, Grainger College of Engineering, University of Illinois Urbana-Champaign. Furthermore, a version of this report has been accepted as an extended abstract for presentation in the Student Research Competition at the \textit{57th ACM Technical Symposium on Computer Science Education} (SIGCSE TS 2026), to be held February 18–21, 2026, in St. Louis, MO, USA.

\balance
\bibliographystyle{ACM-Reference-Format}
\bibliography{sample-base}

@article{Koehler2022,
    author  = {Koehler, A. A. and Meech, S.},
    title   = {Ungrading Learner Participation in a Student-Centered Learning Experience},
    journal = {TechTrends},
    volume  = {66},
    pages   = {78--89},
    year    = {2022}
}

@book{Bloom1956,
    author    = {Bloom, B. S. and Engelhart, M. D. and Furst, E. J. and Hill, W. H. and Krathwohl, D. R.},
    title     = {Taxonomy of Educational Objectives: Cognitive Domain},
    publisher = {Longmans, Green},
    year      = {1956}
}

@article{Lingwall2023,
    author  = {Lingwall, B. N. and Surovek, A. E.},
    title   = {Ungraded Classrooms: A Pattern for Learning in Engineering},
    journal = {ASEE},
    year    = {2023}
}

@inproceedings{Williams2020,
    author    = {Williams, H.},
    title     = {Will Students Engage If There Are No Grades? A Review and Experiment in Ungrading},
    booktitle = {ICERI Proc.},
    year      = {2020},
    pages     = {2575--2581}
}

@article{Cain2022,
    author  = {Cain, J. and Medina, M. and Romanelli, F. and Persky, A.},
    title   = {Deficiencies of Traditional Grading Systems and Recommendations for the Future},
    journal = {Am. J. Pharm. Educ.},
    volume  = {86},
    number  = {7},
    pages   = {8850},
    year    = {2022}
}

@article{Deci2001,
    author  = {Deci, E. L. and Koestner, R. and Ryan, R. M.},
    title   = {Extrinsic Rewards and Intrinsic Motivation in Education: Reconsidered Once Again},
    journal = {Rev. Educ. Res.},
    volume  = {71},
    number  = {1},
    pages   = {1--27},
    year    = {2001}
}

@article{Nathaniel2018,
    author  = {von der Embse, N. and Jester, D. and Roy, D. and Post, J.},
    title   = {Test Anxiety Effects, Predictors, and Correlates: A 30-Year Meta-Analytic Review},
    journal = {J. Affect. Disord.},
    volume  = {227},
    pages   = {483--493},
    year    = {2018}
}

@article{Bensley1992,
    author  = {Bensley, R. and Ellsworth, T.},
    title   = {Bulimic Learning: A Philosophical View of Teaching and Learning},
    journal = {J. Sch. Health},
    volume  = {62},
    pages   = {386--387},
    year    = {1992}
}

@article{Hasinoff2024,
    author  = {Hasinoff, A. A. and Bolyard, W. and DeBay, D. and Dunlap, J. C. and Mosier, A. C. and Pugliano, E.},
    title   = {“Success was Actually Having Learned:” University Student Perceptions of Ungrading},
    journal = {Teach. Learn. Inq.},
    volume  = {12},
    pages   = {1--22},
    year    = {2024}
}

@article{RyanDeci2000,
    author  = {Ryan, R. M. and Deci, E. L.},
    title   = {Self-Determination Theory and the Facilitation of Intrinsic Motivation, Social Development, and Well-Being},
    journal = {Am. Psychol.},
    volume  = {55},
    number  = {1},
    pages   = {68--78},
    year    = {2000}
}

@book{PattPatel,
    author    = {Patt, Y. N. and Patel, S. J.},
    title     = {Introduction to Computing Systems: From Bits \& Gates to C/C++ \& Beyond},
    publisher = {McGraw Hill},
    edition   = {3},
    year      = {2020}
}

@String{Computing = "Computing" }

\end{document}